\documentclass[aps,twocolumn,showpacs,preprintnumbers,amsmath,amssymb,superscriptaddress,floatfix,nofootinbib]{revtex4}
\usepackage{lipsum}
\usepackage{graphicx}
\usepackage{epsfig}
\usepackage{epstopdf}
\usepackage{hyperref}
\usepackage{amsmath}
\usepackage{amsfonts}
\usepackage{amssymb}
\newcommand{\Slash}[1]{\ooalign{\hfil/\hfil\crcr$#1$}}

\begin{document}

\title{$\bar K$ induced formation of the $f_0(980)$ and $a_0(980)$ resonances on proton targets}

\author{Ju-Jun Xie}
\email{xiejujun@impcas.ac.cn} \affiliation{Institute of Modern
Physics, Chinese Academy of Sciences, Lanzhou 730000, China}
\affiliation{State Key Laboratory of Theoretical Physics, Institute
of Theoretical Physics, Chinese Academy of Sciences, Beijing 100190,
China}

\author{Wei-Hong Liang}
\email{liangwh@gxnu.edu.cn}
\affiliation{Department of Physics, Guangxi Normal University,
Guilin 541004, China}

\author{E.~Oset}
\email{oset@ific.uv.es}
\affiliation{Institute of Modern Physics, Chinese Academy of
Sciences, Lanzhou 730000, China} \affiliation{Departamento de
F\'{\i}sica Te\'orica and IFIC, Centro Mixto Universidad de
Valencia-CSIC Institutos de Investigaci\'on de Paterna, Aptdo.
22085, 46071 Valencia, Spain}

\date{\today}

\begin{abstract}

We perform a calculation of the cross section for nine reactions
induced by $\bar K$ scattering on protons. The reactions studied are
$K^-p \to \Lambda \pi^+ \pi^-$, $K^-p \to \Sigma^0 \pi^+ \pi^-$,
$K^-p \to \Lambda \pi^0 \eta$, $K^-p \to \Sigma^0 \pi^0 \eta$, $K^-p
\to \Sigma^+ \pi^- \eta$, $\bar K^0 p \to \Lambda \pi^+ \eta$, $\bar
K^0 p \to \Sigma^0 \pi^+ \eta$, $\bar K^0 p \to \Sigma^+ \pi^+
\pi^-$, $\bar K^0 p \to \Sigma^+ \pi^0 \eta$. We find that in the
reactions producing $\pi^+ \pi^-$ a clear peak for the $f_0(980)$
resonance is found, while no trace of $f_0(500)$ appears. Similarly,
in the cases of $\pi \eta$ production a strong peak is found for the
$a_0(980)$ resonance, with the characteristic strong cusp shape.
Cross sections and invariant mass distributions are evaluated which
should serve, comparing with future data, to test the dynamics of
the chiral unitary approach used for the evaluations and the nature
of these resonances.

\end{abstract}

\maketitle

\section{Introduction}

Kaon beams are becoming a good source for new investigations in
hadron physics. At intermediate energies J-PARC offers good
intensity secondary kaon beams up to about 2
GeV/c~\cite{Sato:2009zze,Kumano:2015gna}. DAPHNE at Frascati
provides low energy kaon beams~\cite{Okada:2010oxa,Bazzi:2013vft}.
Very recently plans have been made for a secondary meson beam
Facility at Jefferson Lab, which includes kaons, both charged and
neutral~\cite{whitebook}. One of the aims is to produce hyperons
($\equiv Y$), which are not as well studied as nucleons or
deltas~\cite{Zhang:2013cua}, and also cascade states, which are even
less known~\cite{Jackson:2013bba,Jackson:2015dva}. In the present
paper we address a different problem using kaon beams, which is the
kaon induced production of the $f_0(980)$ and $a_0(980)$ resonances.
The reactions proposed are $\bar{K} p \to \pi \pi Y$ and $\bar{K} p
\to \pi \eta Y$, which produce the $f_0(980)$ and $a_0(980)$
resonances respectively. These two resonances are the most
emblematic scalar resonances of low energy which have generated an
intense debate as to their nature, as $q \bar q$, tetraquarks, meson
molecules, glueballs, dynamically generated states,
etc.~\cite{Klempt:2007cp}. By now it is commonly accepted that these
mesons are not standard $q \bar q$ states but ``extraordinary"
states~\cite{jaffe}. The coupling of some original $q \bar q$ state
to meson-meson components demanding unitarity has as a consequence
that the meson cloud eats up the original seed becoming the largest
component~\cite{vanBeveren:1986ea,Tornqvist:1995ay,Fariborz:2009cq,Fariborz:2009wf}.
The advent of chiral dynamics in its unitarized form in coupled
channels, the chiral unitary approach, has brought new light into
the subject and the resonances appear from the interaction of
pseudoscalar mesons, usually taken into account by coupled Bethe
Salpeter equations with a kernel, or potential
\cite{npa,kaiser,markushin,juanito} extracted from the chiral
Lagrangians \cite{gasser}, or equivalent methods like the inverse
amplitude method \cite{ramonet,rios}.  A recent review on this issue
makes a detailed comparative study of work done on these issues,
strongly supporting this latter view \cite{sigma}.

The study of $B$ and $D$ decays~\cite{Aaij:2011fx,Muramatsu:2002jp}
has also offered a new valuable source of information on these
states and has stimulated much theoretical
work~\cite{liang,dai,Dedonder:2014xpa,Daub:2015xja,Doring:2013wka,Wang:2015uea,Sekihara:2015iha,Liang:2014ama}.
Yet, little is done in reactions involving baryons, with the
exception of $f_0(980)$ photoproduction, done in
Refs.~\cite{Battaglieri:2008ps,Battaglieri:2009aa}, for which
predictions had been done in Ref.~\cite{Marco:1999nx}, which also
have been addressed theoretically
lately~\cite{daSilva:2013yka,Donnachie:2015jaa}. With this scarce
information, the use of proton targets to produce these states, now
induced by kaons, is bound to be a new good source of information
which should narrow our scope on the nature of these resonances.

One of the outcomes of the chiral unitary theories is that the
$f_0(980)$ couples strongly to $K \bar K$ although it decays into
$\pi \pi$ which is an open channel. On the other hand, the
$a_0(980)$ couples both to $K \bar K$ and to $\pi \eta$, which
becomes the decay channel. The use of kaon beams to produce these
resonances offers one new way in which to test these ideas, since
the original kaon, together with a virtual kaon that will act as a
mediator of the process, will produce the resonances using the
entrance channel to which they couple most strongly. We will study
different processes, having $f_0(980)$ or $a_0(980)$ in the final
state, together with a $\Lambda$ or a $\Sigma$ and we will use both
$K^-$ or $\bar K^0$ to initiate the reaction. In total we study nine
reactions for which we evaluate $d^2\sigma/d M_{\rm inv} d{\rm
cos}(\theta)$ and make predictions for the dependence on the energy
of the beam, the invariant mass of the final two mesons and the
scattering angel, $\theta$.

The contents of the article are organized as follows. In
Sec.~\ref{sec:revisedf0(980)}, we revisit the chiral unitary
approach for the $f_0(980)$ and $a_0(980)$ resonances. In
Sec.~\ref{sec:formalism}, we present the formalism and main
ingredients of the model. In Sec.~\ref{sec:results}, we present our
main results and, finally, in the last section we summarize our
approach and main findings.

\section{The chiral unitary approach for the $f_0(980)$ and $a_0(980)$
resonances} \label{sec:revisedf0(980)}

Following Refs.~\cite{npa,dani}, we start from the coupled channels,
$\pi^+ \pi^-, \pi^0 \pi^0, \pi^0 \eta, \eta \eta, K^+ K^-, K^0 \bar
K^0$, and evaluate the transition potentials from the lowest order
chiral Lagrangians of Ref.~\cite{gasser}. Explicit expressions for
$s$ wave, which we consider here, can be seen in
Refs.~\cite{liang,dai}. Then, by using the on shell factorization of
the Bethe-Salpeter equation in coupled channels~\cite{nsd,ollerulf},
one has in matrix form
\begin{equation}\label{eq:BSeq}
 T=V+VGT; ~~~~~~~T=[1-VG]^{-1} V,
\end{equation}
where $V$ is the transition potential and $G$ the loop function for
two intermediate meson propagators which must be regularized.
Following Ref.~\cite{liang} we take a cut off in three momenta of
600 MeV, demanded when the $\eta \eta$ channel is considered
explicitly. Eq. (\ref{eq:BSeq}) provides the transition $T$ matrix,
$t_{ij}$, from any one to the other channels, and we shall only need
the $t_{K^+K^- \to \pi^+ \pi^-}$, $t_{K^0 \bar K^0 \to \pi^+
\pi^-}$, $t_{K^+K^- \to \pi^0 \eta}$, $t_{K^0 \bar K^0 \to \pi^0
\eta}$ matrix elements. The first two matrix elements contain a pole
associated to the $f_0(980)$, while the latter two contain the pole
of the $a_0(980)$, although this resonance is quite singular and
appears as a big cusp around the $K\bar K$ threshold, both in the
theory as in experiments~\cite{rubin,adams}. The $f_0(980)$ couples
strongly to $K \bar K$ channel with $\pi \pi$ the decay channel, and
the $a_0(980)$ couples strongly to $K \bar K$ and $\pi \eta$
channels.

\section{Formalism} \label{sec:formalism}

\begin{figure}[htbp]
\begin{center}
\includegraphics[scale=0.7]{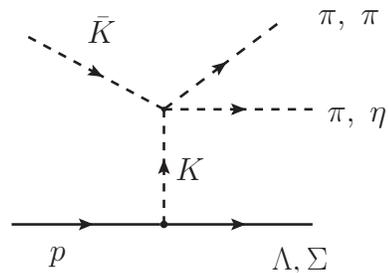}
\caption{Feynman diagram for the $\bar{K} p \to \pi \pi (\pi^0 \eta)
Y$ reaction. \label{fig:FeynmanDiag1}}
\end{center}
\end{figure}

From the perspective that the $f_0(980)$ and $a_0(980)$ resonances
are generated from the meson-meson interaction, the picture for
$f_0(980)$ and $a_0(980)$ anki-kaon induced production proceeds via
the creation of one $K$ by the $\bar{K} p$ initial state in a
primary step and the interaction of the $K$ and $\bar{K}$ generating
the resonances. This is provided by the mechanism depicted in
Fig.~\ref{fig:FeynmanDiag1} by means of a Feynman diagram.

\begin{figure}[htbp]
\begin{center}
\includegraphics[scale=0.8]{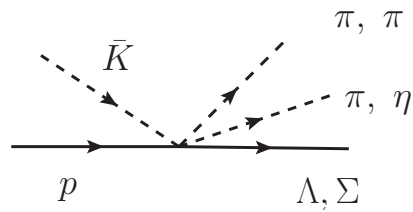}
\caption{Contact term stemming from the Feynman diagram of
Fig.~\ref{fig:FeynmanDiag1} from the off shell part of the $K^+K^-
\to \pi^+\pi^- (\pi^0 \eta)$ transition potential.
\label{fig:FeynmanDiag2}}
\end{center}
\end{figure}

Let us study the $K^- p \to \Lambda (\Sigma^0) \pi^+ \pi^- (\pi^0
\eta)$ as a reference. From this reaction we shall be able to
construct the other five reactions with minimal changes. In this
case, we want to couple the $K^-$ with another $K^+$ to form the
resonances. The first thing one observes is that one of the kaons
(the $K^+$) is necessarily off shell, since neither the $\Lambda$
nor the $\Sigma^0$ can decay into $\bar K p$. Then, in principle one
needs the $K^+ K^- \to \pi^+ \pi^- (\pi^0 \eta)$ amplitude with the
$K^+$ leg off shell, which can be evaluated from the chiral
Lagrangians. Yet, the structure of these Lagrangians is such that
the potential can be written as~\cite{npa}

\begin{eqnarray}\label{eq:potentialV}
V_{K^+K^- \to \pi^+\pi^-} (p_{K^-}, q) &=& V^{\rm on}_{K^+K^- \to
\pi^+\pi^-} (M_{\rm inv}) \nonumber \\
&& + b (q^2-m_{K^+}^2),
\end{eqnarray}
where $p_{K^-}$ and $q$ are the four momenta of $K^-$ and $K^+$
mesons, respectively, while $M_{\rm inv} = \sqrt{(p_{K^-} + q)^2}$
is the invariant mass of the $K^+ K^-$ system. The term with $b$
depends on the representation of the fields taken in the chiral
Lagrangian, while the part of $V^{\rm on}$ does not depend upon this
representation. In this sense the $b$ term is not physical, and
observables cannot depend upon it. The same chiral Lagrangians have
a means to cure this, since the term $b(q^2-m^2_{K^+})$ multiplied
by the $K^+$ propagator of Fig.~\ref{fig:FeynmanDiag1} leads to a
contact term as depicted in Fig.~\ref{fig:FeynmanDiag2}.

However, the chiral Lagrangian for meson baryon~\cite{ecker,ulfrep},
upon expanding on the number of pion fields, contains also contact
terms with the same topology as the one generated from the off shell
part of the amplitude~\cite{alberto} which cancel this latter term.
The result is that one can take just the on shell $K \bar K \to \pi
\pi (\pi \eta)$ amplitude in the diagram of
Fig.~\ref{fig:FeynmanDiag1} and ignore the contact terms stemming
from the meson baryon Lagrangian. These cancelations were observed
before in Ref.~\cite{manolo} in the study of the $\pi N \to \pi \pi
N$ reaction and in Ref.~\cite{carmen} for the study of the pion
cloud contribution to the kaon nucleus optical potential.

The other ingredient that we need for the evaluation of the diagram
of Fig.~\ref{fig:FeynmanDiag1} is the structure of the Yukawa
meson-baryon-baryon vertex. Using chiral Lagrangians~\cite{ecker}
and keeping linear terms in the meson field, the Lagrangian can be
written as
\begin{eqnarray}\label{eq:LagrangianMBB}
\mathcal{L} &=& \frac{D}{2} \langle \bar B \gamma^{\mu} \gamma_5 \{u_{\mu}, B\} \rangle + \frac{F}{2} \langle \bar B \gamma^{\mu} \gamma_5 [u_{\mu}, B] \rangle \nonumber \\
&=& \frac{D+F}{2} \langle \bar B \gamma^{\mu} \gamma_5 u_{\mu} B
\rangle + \frac{D-F}{2} \langle \bar B \gamma^{\mu} \gamma_5 B
u_{\mu}  \rangle,
\end{eqnarray}
where the symbol $<>$ stands for the trace of SU(3). The term linear
in meson field gives
\begin{equation}\label{eq:umu}
u_{\mu}\simeq -\sqrt{2} \frac{\partial_{\mu}\Phi}{f}
\end{equation}
with $f$ the pion decay constant, $f = f_{\pi} = 93$ MeV, and
$\Phi$, $B$ the meson and baryon SU(3) field matrices given by
\begin{eqnarray}
\Phi &=& \left(
\begin{array}{ccc}
\frac{1}{\sqrt{2}} \pi^0 + \frac{1}{\sqrt{6}} \eta & \pi^+ & K^+ \\
\pi^- & - \frac{1}{\sqrt{2}} \pi^0 + \frac{1}{\sqrt{6}} \eta & K^0 \\
K^- & \bar{K}^0 & - \frac{2}{\sqrt{6}} \eta
\end{array}
\right), \\
B &=& \left(
\begin{array}{ccc}
\frac{1}{\sqrt{2}} \Sigma^0 + \frac{1}{\sqrt{6}} \Lambda &
\Sigma^+ & p \\
\Sigma^- & - \frac{1}{\sqrt{2}} \Sigma^0 + \frac{1}{\sqrt{6}} \Lambda & n \\
\Xi^- & \Xi^0 & - \frac{2}{\sqrt{6}} \Lambda
\end{array}
\right).
\end{eqnarray}

We take $F = 0.795$, $D = 0.465$ in this work at the tree level,
consistent with the findings of Ref.~\cite{borasoy}. The explicit
evaluation of the SU(3) matrix elements of
Eq.~\eqref{eq:LagrangianMBB} leads to the following expression

\begin{equation}\label{eq:vertexMBB}
\mathcal{L} \to i \left( \alpha \frac{D+F}{2f} + \beta
\frac{D-F}{2f} \right) \bar u(p',s'_B) \Slash q \gamma_5 u(p,s_B),
\end{equation}
where $u(p,s_B)$ and $\bar u(p',s'_B)$ are the ordinary Dirac
spinors of the initial and final baryons, respectively, and $p$,
$s_B$ and $p'$, $s'_B$ are the four-momenta and spins of the
baryons, while $q = p - p'$ is the four momentum of the meson. The
values of $\alpha$ and $\beta$ are tabulated in
Table~\ref{table:alpha_beta}.

\begin{table}[htbp]
\centering \caption{\small Coefficients for the $\bar{K}NY$
couplings of Eq. (\ref{eq:vertexMBB}).} \vspace{0.5cm}
\begin{tabular}{c|ccc}
         & ~~~$K^-p\to\Lambda$~~~~ & $K^-p\to \Sigma^0$ ~~~~& $K^- n\to \Sigma^-$ \\
         \hline
$\alpha$ & ~~~$-\frac{2}{\sqrt{3}}$ & 0 & 0 \\
$\beta$ & ~~~$\frac{1}{\sqrt{3}}$ & 1 & $\sqrt{2}$ \\
      \hline\hline
         & ~~~$\bar{K}^0 n \to \Lambda$~~~~ & $\bar{K}^0 n \to \Sigma^0$~~~~ &
$\bar{K}^0 p \to \Sigma^+$ \\
         \hline
$\alpha$ & ~~~ $-\frac{2}{\sqrt{3}}$ & 0 & 0 \\
$\beta$ &~~~ $\frac{1}{\sqrt{3}}$ & $-1$ & $\sqrt{2}$
\end{tabular}
\label{table:alpha_beta}
\end{table}

Altogether we can write the amplitude for the diagram of Fig. \ref{fig:FeynmanDiag1} as
\begin{align}\label{eq:amplitude_T}
T= & -i t_{K\bar K \to MM} ~\frac{1}{q^2-m_K^2} \left( \alpha \frac{D+F}{2f} + \beta \frac{D-F}{2f} \right) \nonumber \\
&~ \times \bar u(p',s'_{\Lambda/\Sigma}) \Slash q \gamma_5 u(p,s_p)
F(q^2),
\end{align}
where we have added the customary Yukawa form factor that we take of the form
\begin{equation}\label{eq:FormFactor}
F(q^2) = \frac{\Lambda^2}{\Lambda^2-q^2},
\end{equation}
with typical values of $\Lambda$ of the order of 1 GeV.

The sum and average of $|T|^2$ over final and initial polarization
of the baryons is easily written as
\begin{align}\label{eq:T2}
{\overline {\sum_{s_p}}} \sum_{s'_{\Lambda/\Sigma}} |T_i|^2=&   \left| t^{(i)}_{K\bar K \to MM}\right|^2 \left( \frac{1}{q^2-m_K^2}\right)^2 \times  \nonumber \\
& \frac{(M_p + M')^2}{4M_p M'}  \left[ (M_p - M')^2 -q^2 \right] \times  \nonumber \\
&\left( \alpha_i \frac{D+F}{2f} + \beta_i \frac{D-F}{2f} \right)^2
F^2(q^2),
\end{align}
where $M_p, M'$ are the masses of the proton and the final baryon
($\Lambda$ or $\Sigma$). The subindex $i$ stands for different
reactions.

We can write $q^2$ in terms of the variables of the external
particles and have
\begin{equation}\label{eq:q2}
q^2 = M_p^2 + M'^2-2E E' + 2|\vec{p}| |\vec{p'}| \cos \theta,
\end{equation}
where $\vec{p},\vec{p'}$ and $E, E'$ are the momenta and energies of
the proton and the final baryon, and $\theta$ is the angle between
the direction of the initial and final baryon, all of them in the
global center of mass frame (CM). The $\vec{p},\vec{p'}$ and $E, E'$
have the form as
\begin{eqnarray}
|\vec{p}| &=& \frac{\lambda^{1/2}(s, m_{\bar
K}^2,M_p^2)}{2\sqrt{s}}, \\
|\vec{p'}| &=& \frac{\lambda^{1/2}(s, M_{\rm
inv}^2,M'^2)}{2\sqrt{s}}, \\
E &=& \sqrt{M^2_p + |\vec{p}|^2}, \\
E' &=& \sqrt{M'^2 + |\vec{p'}|^2},
\end{eqnarray}
where $s$ is the invariant mass square of the $\bar K p$ system and
$\lambda$ is the K\"allen function with $\lambda(x,y,z) = (x-y-z)^2
- 4yz$.

We can write the differential cross section as
\begin{equation}\label{eq:CrossSection}
\frac{{\rm d}^2 \sigma}{{\rm d} M_{\rm inv} {\rm d}\cos \theta} =
\frac{M_p M'}{32
\pi^3}\frac{|\vec{p'}|}{|\vec{p}|}\frac{|\vec{\tilde{p}}|}{s}
{\overline {\sum_{s_p}}} \sum_{s'_{\Lambda/\Sigma}} |T|^2 ,
\end{equation}
with $|\vec {\tilde p}|$ the momentum of one of the mesons in the
frame where the two final mesons are at rest,
\begin{equation}\label{eq:tilde_p1}
|\vec{\tilde p}| = \frac{\lambda^{1/2}(M_{\rm inv}^2,
m_1^2,m_2^2)}{2M_{\rm inv}},
\end{equation}
where $M_{\rm inv}$ is the invariant mass of the two mesons system,
and $m_1$ and $m_2$ are the masses of the two mesons, respectively.
Note that the $K \bar K \to MM$ scattering amplitudes $t_{K\bar K
\to MM}$ depend on $M_{\rm inv}$ only.

We want to study nine reactions
\begin{align}\label{eq:9Reactions}
& K^-p \to \Lambda \pi^+ \pi^-, ~~ K^-p \to \Sigma^0 \pi^+ \pi^-, ~~K^-p \to \Lambda \pi^0 \eta, \nonumber \\
& K^-p \to \Sigma^0 \pi^0 \eta, ~~ K^-p \to \Sigma^+ \pi^- \eta, ~~\bar K^0p \to \Lambda \pi^+ \eta, \\
& \bar K^0 p \to \Sigma^0 \pi^+ \eta,~~\bar K^0 p \to \Sigma^+ \pi^+ \pi^-,~~\bar K^0 p \to \Sigma^+ \pi^0 \eta. \nonumber
\end{align}
The Yukawa vertices for $K B B $ are summarized in
Table~\ref{table:alpha_beta}. The $K \bar K \to MM$ amplitudes are
discussed above. However, only the $I_3 = 0$ components are studied
there, corresponding to zero charge. We have three cases with $\pi
\eta$ where the charge is non zero, $K^-p \to \Sigma^+ \pi^- \eta$,
$\bar K^0 p \to \Lambda \pi^+ \eta$ and $\bar K^0 p \to \Sigma^0
\pi^+ \eta$. We can easily relate the $K \bar K \to \pi \eta$
amplitudes to the $K^+ K^- \to \pi^0 \eta$, which is evaluated in
the case of zero charge, using isospin symmetry. Indeed, recalling
the phases $|K^->= |1/2,-1/2>$, $|\pi^+>=-|1,1>$, we can write in
terms of the total isospin
\begin{align}\label{eq:isospin}
&~|K^+K^-\rangle = - \frac{1}{\sqrt{2}} |1,0\rangle - \frac{1}{\sqrt{2}} |0,0\rangle,   \nonumber \\
&~ |K^0 K^-\rangle = -|1,-1\rangle, ~~ |K^+ \bar K^0\rangle =
|1,1\rangle, \nonumber \\
&~ |\pi^+ \eta\rangle = -|1,1\rangle,~~~|\pi^- \eta\rangle =
|1,-1\rangle,
\end{align}
and then we find
\begin{align}\label{eq:t_related}
t_{K^+K^- \to \pi^0 \eta}&= -\frac{1}{\sqrt{2}}~t^{I=1}_{K\bar K \to \pi \eta},\nonumber \\
t_{K^0K^- \to \pi^- \eta}&= \sqrt{2}~t_{K^+K^- \to \pi^0 \eta},\\
t_{K^+ \bar K^0 \to \pi^+ \eta}&= \sqrt{2}~t_{K^+K^- \to \pi^0 \eta}.\nonumber
\end{align}
With these ingredients we will use Eq. (\ref{eq:CrossSection}) to
evaluate the cross section in each case, and all we must do is
change the $t_{k \bar K, MM}$ in each case and the values of
$\alpha$ and $\beta$. These magnitudes are summarized in
Table~\ref{table:t_alpha_beta}.

\begin{table}[htbp]
  \renewcommand{\arraystretch}{1.5}
  \setlength{\tabcolsep}{0.2cm}
  \centering
  \caption{Matrices $t_{K\bar K \to MM}$, $\alpha$, $\beta$ used in each reaction and resonance obtained.}
  \begin{ruledtabular}
  \begin{tabular}{llccc}
    Reaction         & $t_{K\bar K \to MM}$ & $\alpha$ &  $\beta$  & Resonance\\
  \hline
  $K^-p \to \Lambda \pi^+ \pi^-$ & $t_{K^+ K^- \to \pi^+ \pi^-}$ & $-\frac{2}{\sqrt{3}}$ & $\frac{1}{\sqrt{3}}$ & $f_0(980)$  \\
  $K^-p \to \Sigma^0 \pi^+ \pi^-$ & $t_{K^+ K^- \to \pi^+ \pi^-}$ & $0$ & $1$ & $f_0(980)$  \\
  $K^-p \to \Lambda \pi^0 \eta$ & $t_{K^+ K^- \to \pi^0 \eta}$ & $-\frac{2}{\sqrt{3}}$ & $\frac{1}{\sqrt{3}}$ & $a_0(980)$  \\
  $K^-p \to \Sigma^0 \pi^0 \eta$ & $t_{K^+ K^- \to \pi^0 \eta}$ & $0$ & $1$ & $a_0(980)$  \\
  $K^-p \to \Sigma^+ \pi^- \eta$ & $\sqrt{2}~t_{K^+ K^- \to \pi^0 \eta}$ & $0$ & $\sqrt{2}$ & $a_0(980)$  \\
  $\bar K^0 p \to \Lambda \pi^+ \eta$ & $\sqrt{2}~t_{K^+ K^- \to \pi^0 \eta}$ & $-\frac{2}{\sqrt{3}}$ & $\frac{1}{\sqrt{3}}$ & $a_0(980)$  \\
  $\bar K^0 p \to \Sigma^0 \pi^+ \eta$ & $\sqrt{2}~t_{K^+ K^- \to \pi^0 \eta}$ & $0$ & $1$ & $a_0(980)$  \\
  $\bar K^0 p \to \Sigma^+ \pi^+ \pi^-$ & $t_{K^0 \bar K^0 \to \pi^+ \pi^-}$ & $0$ & $\sqrt{2}$ & $f_0(980)$  \\
   $\bar K^0 p \to \Sigma^+ \pi^0 \eta$ & $t_{K^0 \bar K^0 \to \pi^0 \eta}$ & $0$ & $\sqrt{2}$ & $a_0(980)$  \\
  \end{tabular}
  \end{ruledtabular}
  \label{table:t_alpha_beta}
\end{table}

\section{Results} \label{sec:results}

\begin{figure}[htbp]
\begin{center}
\includegraphics[scale=0.45]{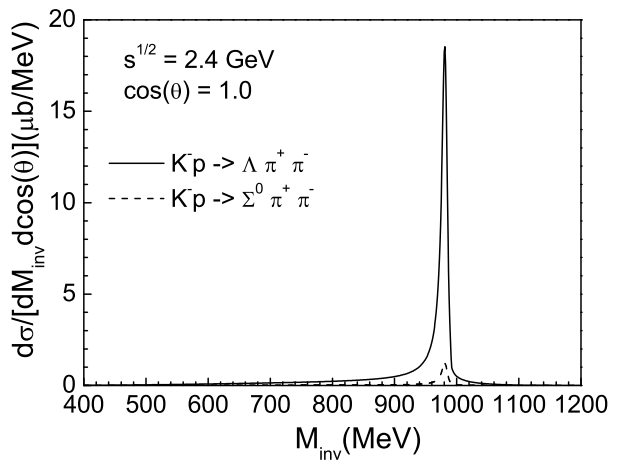}
\caption{Theoretical predictions for $S$ wave $\pi^+ \pi^-$ mass
distributions for $K^- p \to \Lambda (\Sigma^0) \pi^+ \pi^-$
reactions at $\sqrt{s} = 2.4$ GeV and ${\rm cos}(\theta) = 1$.}
\label{Fig:dsigdm-kmp-f0}
\end{center}
\end{figure}

We have a dependence of the cross section in the energy, $M_{\rm
inv}$, and scattering angle $\theta$ given by Eq.~(\ref{eq:q2}). We
first evaluate the cross section for $\theta=0$, in the forward
direction. In Fig.~\ref{Fig:dsigdm-kmp-f0}, we show the numerical
results of ${\rm d} \sigma / {\rm d} M_ {\rm inv} {\rm d} \cos
\theta$ for $\cos (\theta) =1$ as a function of $M_{\rm inv}$ of the
$\pi^+ \pi^-$ for $K^- p \to \Lambda (\Sigma^0) \pi^+ \pi^-$
reactions. We have chosen $\sqrt s= 2.4 ~{\rm GeV}$, corresponding
to the $K^-$ momentum $p_{K^-} = 2.42 ~{\rm GeV}$ in the laboratory
frame.~\footnote{In the laboratory frame, $s = m^2_{\bar K} + m^2_p
+ 2m_p \sqrt{m^2_{\bar K} + p^2_{\bar K}}$.} One can see that there
is a clear peak around $M_{\rm inv} = 980$ MeV which is the signal
for the $f_0(980)$ resonance that was produced by the initial $K^+
K^-$ coupled channel interactions and decaying into $\pi^+ \pi^-$
channel. On the other hand, the magnitude of the cross section for
$\Lambda$ production is of the order of 10 times larger than for
$\Sigma^0$ production, because the coupling of $KN\Lambda$ is
stronger than the $KN\Sigma$ coupling.

\begin{figure}[htbp]
\begin{center}
\includegraphics[scale=0.45]{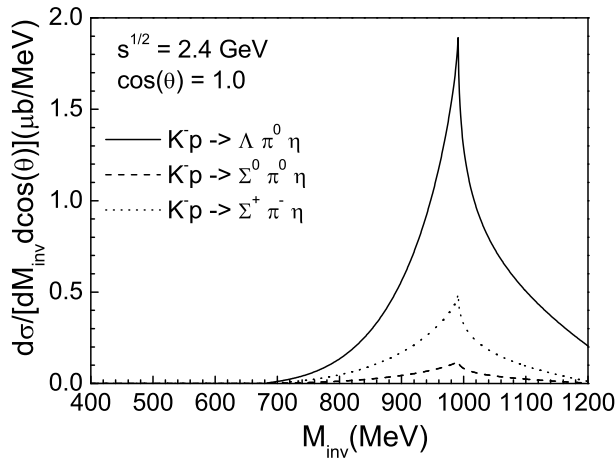}
\caption{Theoretical predictions for $S$ wave $\pi \eta$ mass
distributions for $K^- p \to \Lambda (\Sigma^0) \pi^0 \eta$ and $K^-
p \to \Sigma^+ \pi^- \eta$ reactions at $\sqrt{s} = 2.4$ GeV and
${\rm cos}(\theta) = 1$.} \label{Fig:dsigdm-kmp-a0}
\end{center}
\end{figure}

In Fig.~\ref{Fig:dsigdm-kmp-a0}, we show the numerical results of
${\rm d} \sigma / {\rm d} M_ {\rm inv} {\rm d} \cos \theta$ for
$\cos (\theta) =1$ as a function of $M_{\rm inv}$ of the $\pi \eta$
for $K^- p \to \Lambda (\Sigma^0) \pi^0 \eta$ and $K^- p \to
\Sigma^+ \pi^- \eta$ reactions. In this case we see also a clear
peak/cusp around $M_{\rm inv} = 980$ MeV which corresponds to the
$a_0(980)$ state.

\begin{figure}[htbp]
\begin{center}
\includegraphics[scale=0.45]{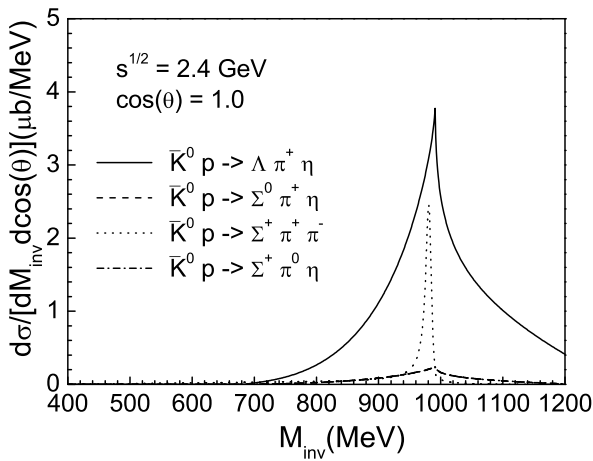}
\caption{Theoretical predictions for $S$ wave $\pi \eta$ and $\pi^+
\pi^-$ mass distributions for $\bar{K}^0 p \to \Lambda (\Sigma^0)
\pi^+ \eta$ and $K^- p \to \Sigma^+ \pi^0 \eta (\pi^+ \pi^-)$
reactions at $\sqrt{s} = 2.4$ GeV and ${\rm cos}(\theta) = 1$.}
\label{Fig:dsigdm-k0barp}
\end{center}
\end{figure}

Similarly, we show our results for $\bar{K}^0 p$ reactions in
Fig.~\ref{Fig:dsigdm-k0barp}. One can see again the clear peaks for
$a_0(980)$ and $f_0(980)$ resonances around $M_{\rm inv} = 980$ MeV.

In all the reactions mentioned above, we observe clear peaks for the
$f_0(980)$ in the case of the $\pi^+ \pi^-$ production or for the
$a_0(980)$ in the case of $\pi \eta$ production. It is remarkable
that in the case of the $f_0(980)$ production there is no trace of
the $f_0(500)$ ($\sigma$) production. This is reminiscent of what
happens in $B^0_s \to J/\psi \pi^+ \pi^-$, where a clear peak is
seen for the $f_0(980)$ but no trace is observed of the $f_0(500)$
\cite{Aaij:2011fx}. The chiral unitary approach of Ref.~\cite{liang}
offered an explanation for this fact. Indeed, in this reaction at
the quark level one produces $c \bar c$, that makes the $J/\psi$,
and a $s \bar s$ pair. This pair hadronizes into two mesons which
are not $\pi \pi$, but mostly $K \bar K$ or $\eta \eta$. Then these
particles undergo final state interaction producing the resonances.
However, the $K \bar K$ couples strongly to the $f_0(980)$ resonance
and very weakly to the $f_0(500)$, and this explains the observed
features. In this case we have the $K  \bar{K}$ producing the
resonances and, similarly, we find a production of the $f_0(980)$
and not of the $f_0(500)$.

The reactions with $\pi \eta$ in the final state produce the
$a_0(980)$ resonance. It is interesting to observe the shape. It is
much as a cusp around the $K \bar K$ threshold, but with a large
strength. As we remarked earlier, this feature is common to all
reactions where the $a_0(980)$ is produced with good
statistics~\cite{rubin,adams}.

\begin{figure}[htbp]
\begin{center}
\includegraphics[scale=0.45]{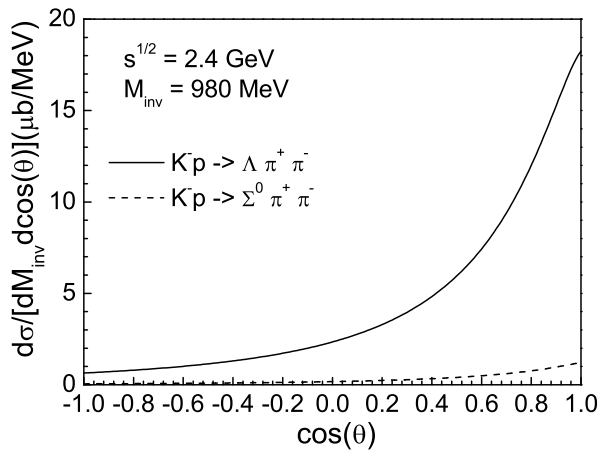}
\caption{Theoretical predictions for ${\rm d} \sigma / {\rm d} M_
{\rm inv} {\rm d} \cos \theta$ as a function of ${\rm cos}(\theta)$
for $K^- p \to \Lambda (\Sigma^0) \pi^+ \pi^-$ reactions at
$\sqrt{s} = 2.4$ GeV and $M_{\rm inv} = 980$ MeV.}
\label{Fig:dsigdm-kmp-f0-costheta}
\end{center}
\end{figure}

\begin{figure}[htbp]
\begin{center}
\includegraphics[scale=0.45]{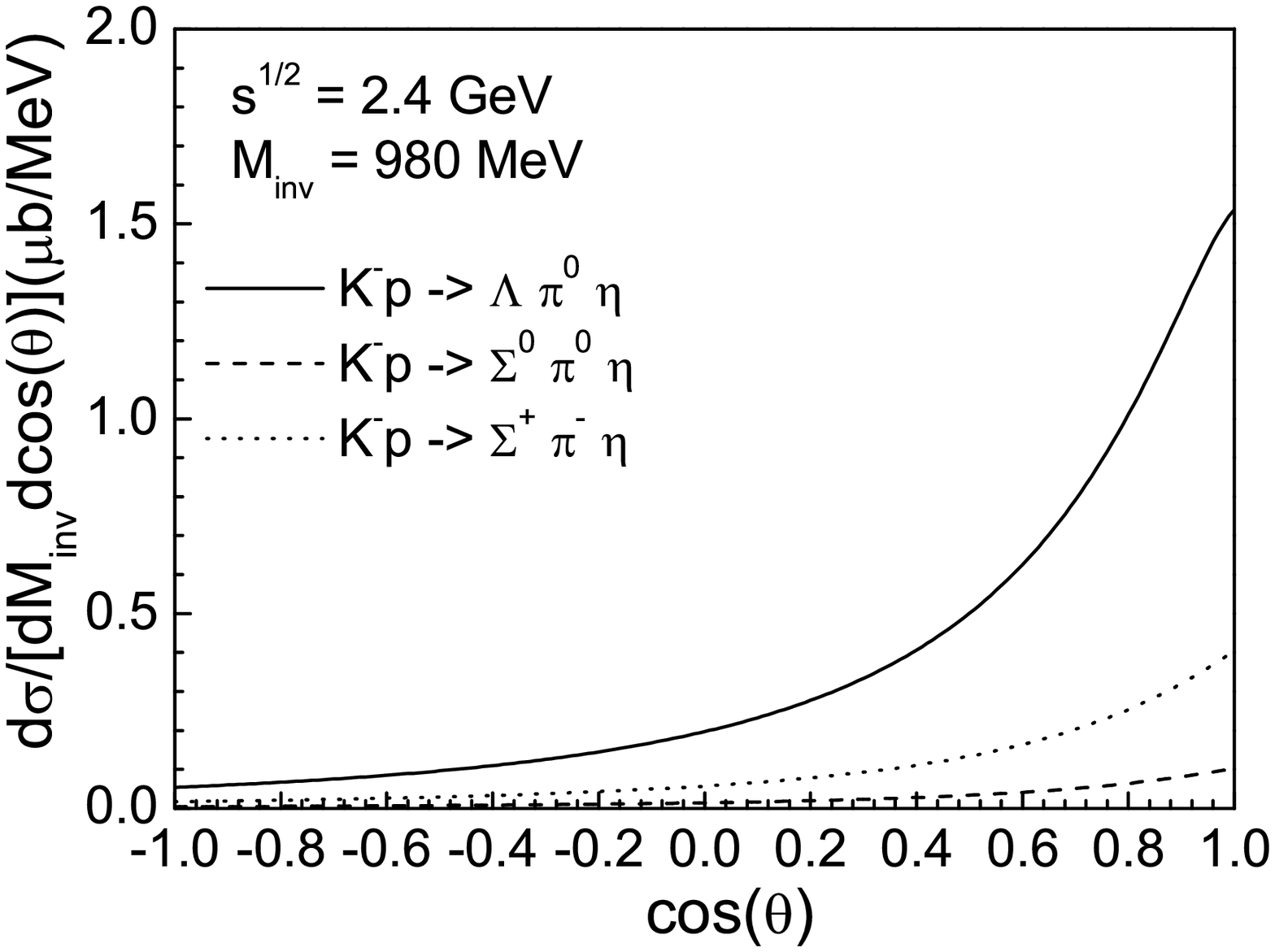}
\caption{Theoretical predictions for ${\rm d} \sigma / {\rm d} M_
{\rm inv} {\rm d} \cos \theta$ as a function of ${\rm cos}(\theta)$
for $K^- p \to \Lambda (\Sigma^0) \pi^0 \eta$ and $K^- p \to
\Sigma^+ \pi^- \eta$ reactions at $\sqrt{s} = 2.4$ GeV and $M_{\rm
inv} = 980$ MeV.} \label{Fig:dsigdm-kmp-a0-costheta}
\end{center}
\end{figure}

\begin{figure}[htbp]
\begin{center}
\includegraphics[scale=0.45]{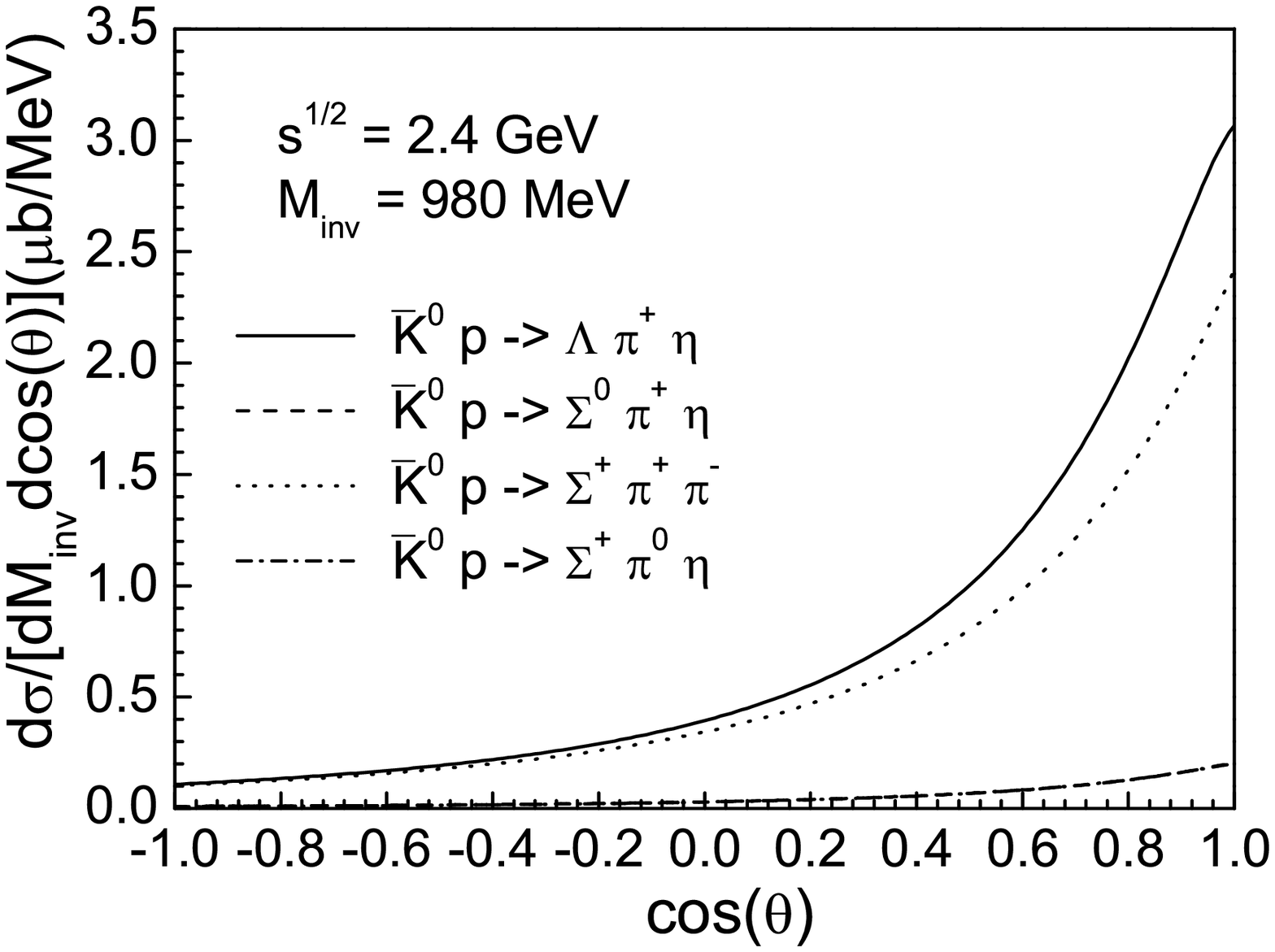}
\caption{Theoretical predictions for ${\rm d} \sigma / {\rm d} M_
{\rm inv} {\rm d} \cos \theta$ as a function of ${\rm cos}(\theta)$
for $\bar{K}^0 p \to \Lambda (\Sigma^0) \pi^+ \eta$ and $\bar{K}^0 p
\to \Sigma^+ \pi^0 \eta (\pi^+ \pi^-)$ reactions at $\sqrt{s} = 2.4$
GeV and $M_{\rm inv} = 980$ MeV.} \label{Fig:dsigdm-k0barp-costheta}
\end{center}
\end{figure}

Furthermore, in Figs.~\ref{Fig:dsigdm-kmp-f0-costheta} to
\ref{Fig:dsigdm-k0barp-costheta} we show the results for ${\rm d}
\sigma / {\rm d} M_ {\rm inv} {\rm d} \cos \theta$ for the $\bar{K}
p$ reactions at the peak of the invariant mass, $f_0(980)$,
$a_0(980)$ respectively, as a function of $\cos \theta$. Because we
considered only the contributions from the $t$ channel $K$ exchange,
the reactions peak forward and one can see a fall down of about of a
factor 10 in the cross section from forward to backward angles,
where contributions from $s$ and $u$ channels could be dominant.

\begin{figure}[htbp]
\begin{center}
\includegraphics[scale=0.45]{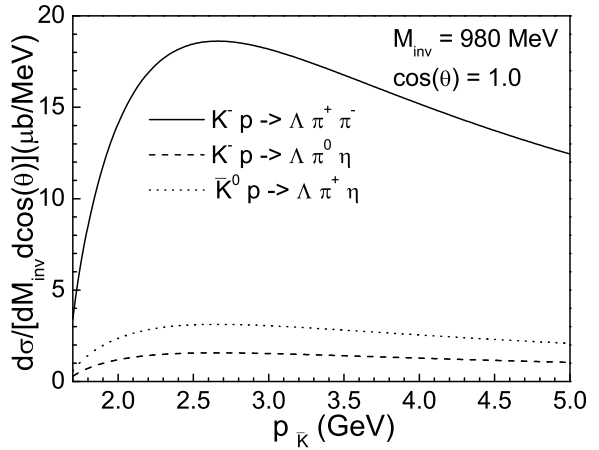}
\caption{Theoretical predictions for ${\rm d} \sigma / {\rm d} M_
{\rm inv} {\rm d} \cos \theta$ as a function of $p_{\bar K}$ for
$K^- p \to \Lambda \pi^+ \pi^- (\pi^0 \eta)$ and $\bar{K}^0 p \to
\Lambda \pi^+ \eta$ reactions at ${\rm cos} \theta = 1$ and $M_{\rm
inv} = 980$ MeV.} \label{Fig:dsigdm-lambda-plab}
\end{center}
\end{figure}

Finally, we now fix $M_{\rm inv} = 980$ MeV at the peak of the
resonance and $\cos \theta =1$ and look at the dependence of the
cross section with the energy of the $\bar K $ beam. Because the
$\Lambda$ production is larger than the $\Sigma$ production, we show
only the results for the $\Lambda$ production in
Fig.~\ref{Fig:dsigdm-lambda-plab}. We observe that the cross section
grows fast from the reaction threshold and reaches a peak around
$p_{\bar K} = 2.5$ GeV.

\section{Conclusions} \label{sec:summary}

In this work, we study the production of $f_0(980)$ and $a_0(980)$
resonances in the $\bar{K} p$ reaction with the picture that these
two resonances are dynamically generated within the coupled
pseudoscalar-pseudoscalar channels interaction in $I = 0$ and $1$,
respectively. This is the first evaluation of the cross section for
these reactions. In the cases of $\pi^+ \pi^-$ production we find a
neat peak for the $f_0(980)$ production and no production of the
$f_0(500)$. This feature is associated to the fact that the
resonance is created from $K \bar K$ and the $f_0(980)$ has a strong
coupling $K \bar K$ while the $f_0(500)$ has a very small coupling
to this component. Thus, in spite of the fact that the $f_0(980)$ is
observed in the $\pi^+ \pi^-$, to which the $f_0(500)$ couples
strongly, one finishes with a negligible signal for $f_0(500)$ in
this reaction. This feature is also observed in the $B_s \to J/\psi
\pi^+ \pi^-$ reaction and we find a natural explanation of both
reactions within the chiral unitary approach to the nature of these
resonances. It would be good to have the reactions proposed
implemented in actual experiments to narrow the scope on possible
interpretations of the nature of these resonances. Some alternative
explanations for the features observed in the $B_s \to J/\psi \pi^+
\pi^-$ reaction are given for instance in Ref.~\cite{sheldon} and
would be good to see what these pictures would predict for the
reactions studied here.

The reactions with the $\pi \eta$ production give rise also to a
clear peak corresponding to the $a_0(980)$. This resonance appears
as border line in the chiral unitary approach, corresponding to a
state slightly unbound, or barely bound. The fact is that is shows
up clearly in form of a strong cusp around the $K \bar K$ threshold,
and this feature is observed in recent experiments with large
statistics. It would be good to see what happens when the experiment
is done. We should also note that our theoretical approach provides
the absolute strength for both the $f_0(980)$ and $a_0(980)$
production and this is also a consequence of the theoretical
framework that generates dynamically these two resonances.

We have assumed a $t$-channel dominance, based on the strong
coupling of the resonances to $K \bar K$. This has as a consequence
that the nine reactions that we have studied have a definite weight,
the largest differences coming from the Yukawa MBB couplings which
are well known. Comparison of the strength of these reactions could
serve to assert the dominance of the production model that we have
assumed.

\section*{Acknowledgments}

One of us, E. O., wishes to acknowledge support from the Chinese
Academy of Science in the Program of Visiting Professorship for
Senior International Scientists (Grant No. 2013T2J0012). This work
is partly supported by the Spanish Ministerio de Economia y
Competitividad and European FEDER funds under the contract number
FIS2011-28853-C02-01 and FIS2011-28853-C02-02, and the Generalitat
Valenciana in the program Prometeo II-2014/068. We acknowledge the
support of the European Community-Research Infrastructure
Integrating Activity Study of Strongly Interacting Matter (acronym
HadronPhysics3, Grant Agreement n. 283286) under the Seventh
Framework Programme of EU. This work is also partly supported by the
National Natural Science Foundation of China under Grant Nos.
11165005, 11565007 and 11475227. This work is also supported by the
Open Project Program of State Key Laboratory of Theoretical Physics,
Institute of Theoretical Physics, Chinese Academy of Sciences, China
(No.Y5KF151CJ1).

\bibliographystyle{plain}

\end{document}